\documentstyle[12pt]{article}
\begin{document}
\author{
Santi Prestipino$^{a,b}$\cite{aff1}, Eduardo A. Jagla$^{c}$\cite{aff2},
and Erio Tosatti$^{a,b,c}$\cite{aff3} \\
{\it a) Istituto Nazionale per la Fisica della Materia (INFM)}\\
{\it b) International School for Advanced Studies, Trieste, Italy}\\
{\it c) The Abdus Salam International Centre for Theoretical Physics,}\\
{\it Trieste, Italy}
}
\title{Can one have preroughening \\ of vicinal surfaces?}
\date{\today}
\maketitle
\begin{abstract}
We discuss the possibility that, besides roughening, a vicinal surface
could display preroughening (PR), and consider the possible mechanisms
for its promotion. Within the framework of a terrace-step-kink model,
it turns out that a PR transition is possible, and could be induced 
by a short-range repulsion between parallel kinks along the same step
or on adjacent steps, or even by some kind of extended range step-step
repulsion. We discuss the possible relevance of this phenomenon to the
anomalous roughening behaviour recently reported for Ag(115).

\vspace{3mm}
\end{abstract}

\section{Introduction}

The identification of a new thermodynamic surface phase is always of
fundamental significance. Vicinal surfaces of metals, consisting of a 
regular array of low-index terraces separated by parallel steps, are 
of special interest because of their ubiquity, due to {\em e.g.} small
miscut angles, and because of the expected rich interplay between steps
and kinks. A vicinal surface typically has only two states: flat and
rough, with a phase transition from flat to rough at a much lower
temperature than that of a low-index face. The transition is generated
by the proliferation of kinks in the steps, as first discussed by
Villain {\em et al.}\cite{Villain} At roughening, steps develop kinks
and start to meander freely, giving rise to divergent fluctuations
in the position of the surface along the normal. At these relatively
low temperatures, very neat STM observations of roughening are possible,
as recently demonstrated for Ag(115) by Hoogeman {\em et al.}\cite{STM} 

A more complex and interesting roughening behaviour could be anticipated
if interactions between the kinks were important. Elasticity theory
predicts a repulsive interaction between two kinks in the same direction
along a step (parallel kinks) which asymptotically should decay as
$r^{-3}$.\cite{Marchenko} If this repulsion were strong enough, a new
vicinal disordered flat (VDOF) phase, with long-range antiparallel order
of kinks within a step, could perhaps be stabilized, in analogy with the
stabilization on a low-index surface of a standard DOF phase\cite{MdN89}
by elastic parallel step-step repulsion.\cite{Santi95} Similarly, 
the effect of parallel kink-kink repulsion should be to shift vicinal
roughening to a higher temperature, and this may uncover a vicinal
preroughening (VPR) transition which, involving proliferation
of correlated left-right kinks, remains unaffected.

We have identified a simplified model of a vicinal surface where 
precisely this scenario is realized. While we believe that our
findings have a broader significance, we have specialized our
modeling to mimick the situation of Ag(115), where we speculate
VPR might take place.

\section{Model and Method}

So long as temperature is far below melting, a good description of
vicinal surface roughening is given by the terrace-step-kink (TSK)
model.\cite{Lapujoulade} It ignores the possibility of adatoms on
terraces, so that the step positions are the only statistical variables
in the problem. If a solid-on-solid condition is assumed all over the
step length, then the position of a step with respect to $T=0$ can be
encoded by a variable $u_m(y)$ (where $m$ is the step label and $y$ is
an abscissa along the step). The TSK model further assumes a repulsive
interaction between neighbouring steps and an energy cost for forming
kinks on primary steps. In the celebrated model by Villain, Grempel, and
Lapujoulade (VGL),\cite{Villain} $u_m(y)$ is an integer variable and only
unitary kinks are permitted (the energy cost for each being $W_1>0$).
Moreover, an energy $U_1>0$ is to be paid when
$\Delta u=u_{m+1}(y)-u_m(y)=-1$, nothing for $\Delta u>-1$, while
$\Delta u<-1$ is not allowed. The VGL model is exactly solvable in the
highly anisotropic, $W_1/U_1\rightarrow +\infty$ limit.\cite{Villain}
In particular, a Kosterlitz-Thouless phase transition is predicted to
occur at a temperature $T_{\rm R}$ given by $U_1/\left( 2k_BT_{\rm R}\right)
\exp\left\{W_1/\left( k_BT_{\rm R}\right) \right\}=1$. This transition
describes surface roughening since the correlation function
$G(m)=\left\langle\left( u_m(y)-u_0(y)\right) ^2\right\rangle$ is 
finite as $m$ goes to infinity when $T<T_{\rm R}$, while it diverges
logarithmically when $T>T_{\rm R}$. 

We have modified the VGL Hamiltonian to include interactions between
the kinks. These interactions generally arise from the mutual 
interference between the elastic strain fields determined by the 
individual kinks. In particular, we allow for a short-range 
repulsion ($W_2$) between
parallel kinks in the same step, and also between parallel kinks in
neighbouring steps ($U_2$). Moreover, the two-sublattice structure
of fcc(115) is also kept into account (see Fig.\,1). 

With these ingredients, we have applied standard Metropolis Monte Carlo
or alternatively, mapping onto a 1D quantum spin chain to study the 
phase diagram as a function of temperature. The Monte Carlo lattices
employed comprise an increasing number of steps, up to $N_x=24$,
while the number $N_y$ of step sites along the hard $y$ direction
is taken 60 times larger.
We monitor usual quantities like kink-kink correlations, heat capacity,
VDOF order parameter $P=\left\langle (-1)^{u_m(y)}\right\rangle$, and the
average square lateral excursion of a step $\delta u^2=\left\langle\left(
u_m(y)-\overline{u_m}\right) ^2\right\rangle$.
Strong antiparallel VDOF correlations between consecutive kinks in a step
will result into a vanishing $P$ in the thermodynamic limit, with a finite
$\delta u^2$ infinite-size extrapolation. In the ordered flat phase,
conversely, $P$ is finite, whereas in the disordered rough phase, 
$\delta u^2$ extrapolates to infinity.

\section{Results}

First, we prove the existence of a VDOF phase in the phase diagram of
our model vicinal by Monte Carlo. To this end, we observe that the most
favourable situation for the VDOF is when $U_2$ and $W_2$ are both infinite.
We take $W_1=20U_1$ and vary $\beta U_1\equiv U_1/\left( k_BT\right) $
from 0 to 0.18. We find that the rough phase is altogether absent in this
case and the high-temperature phase of the model is VDOF. In Fig.\,2, a
small part of the vicinal surface is shown at infinite temperature. A glance
at this picture reveals the general structure of the VDOF phase in our model:
kinks are very numerous along the steps but the correlation between two
consecutive kinks is strictly antiparallel, so that the overall surface
is flat and its slope well-defined. The VDOF state in Fig.\,2 is nearly
ideal: each step strictly meanders between two positions with equal
probability, whence $\delta u^2\simeq 0.25$. On decreasing temperature,
$\delta u^2$ shows a peak, centered at $\beta U_1\simeq 0.08$, which is
size-dependent and allows to locate VPR\cite{Santi95}
($\delta u^2=0.473,0.580,0.618$
for $N_x=8,16,24$, when $\beta U_1=0.08$; a $K\ln N$ fit of the data
for $N_x=16$ and 24 yields $K\approx 0.09$). At the VPR temperature,
$\delta u^2$ grows logarithmically as a function of the surface size
and the vicinal is thus rough at an isolated temperature point,
while being flat both below and above. We conclude that if the short-range
repulsion between parallel kinks is strong enough, then generally 
three different surface phases will be thermodinamically stable, i.e.,
a low-temperature ordered flat phase, an intermediate VDOF phase,
and a rough phase at sufficiently high $T$. A weaker repulsion between
the kinks, on the other hand, may not be sufficient to stabilize the
VDOF phase, and in that case the standard flat-rough transition is
recovered, as in the straight VGL model.

Next, we consider the phase diagram for $W_2=U_2=0$, but with an additional
interaction similar to $U_1$ assumed between {\em second}-neighbour steps
too. Precisely, we assume an energy cost $U_1^{'}$ whenever two second
neighbouring steps get to the minimal distance $3a$, where $a$ is the atomic
diameter. This discourages meandering of the steps without affecting their
kink-antikink zigzagging between two positions only. By mapping onto a 1D
quantum model,\cite{Villain,KK} we confirm that a VDOF phase can be generated
by this second-neighbour step repulsion. While the precise value $U_1^{'}$
assumed is of course unrealistic since the repulsive force between two
distant steps decay as $r^{-2}$ with distance,\cite{Marchenko} this
represents an interesting alternative mechanism to parallel kink repulsion
to generate a VDOF phase.

\section{Is there a VDOF phase on Ag(115)?}

Having satisfied ourselves that VPR transitions and VDOF phases exist
in a suitable vicinal model, it becomes of course of interest to ask
whether they might actually be realized in a real system.
Hoogeman {\em et al.} have carried out the first really extensive STM
study of a vicinal surface, chosen to be Ag(115).\cite{STM} Their
analysis indicates that, in the temperature range 400-500 K, 
the correlation function
$G(m)$ of this surface is well represented by a law of the type
$-K\ln\left( m^{-2}+X^{-2}\right) +C$, which people use for standard
surface roughening. Strikingly, however, $X^{-1}$ extrapolates to
zero at about 440 K, while $K$ crosses the universal roughening value
$1/\pi^2$ only at the much higher temperature of 490 K. In regular
roughening, these two temperatures ought to coincide. It is tempting
to hypothesize that the divergence of $X$ could now indicate VPR at
440 K, so that a VDOF phase would be realized on Ag(115) in the
temperature window 440-490 K.

What could be a realistic set of parameters to represent Ag(115)?
Within the framework of the TSK model with interacting kinks, the problem
of fitting the parameters to a realistic vicinal is of no simple solution.
For Ag(115), Hoogeman {\em et al.} have shown that the kink formation energy
and the step repulsion can be extracted from STM data, thus providing values
for $W_1$ and $U_1$.\cite{nota}
Moreover, even without a systematic study of kink-kink interactions,
a value of 260 K was reported for the repulsion between kinks
in neighbouring steps, which we take as the value of $U_2$.
Finally, we have no information on $W_2$, but we believe it should be
smaller than $U_2$, and thus tentatively set it to zero.

In Fig.\,3, we show simulation data for three sizes, $N_x=12,18,$ and
24. Even though the statistics is insufficient, both these data and the
behavior of the $u$ values during the run are suggestive of a VPR
transition at about 380 K, while roughening would not occur before 390 K.
Alternatively, the VDOF state which is occasionally observed during the
simulation is in fact only metastable, and a single roughening transition
actually occurs at about 375 K, as suggested by the thermal behaviour
of $G(m)$.
Summing up at this stage, we find that with this first choice of parameters
there are hints of VPR and of a narrow DOF region. However, the evidence
is quantitatively inconclusive. Further work is now in progress, using a
different set of parameters. In particular, it is expected that
the effect of a positive $W_2$ should be to shift roughening
(and to a smaller extent also preroughening) further up in temperature,
which goes qualitatively in the right direction.
A more detailed account of ongoing calculations will be presented
elsewhere.\cite{KK}

\section{Conclusions}

In conclusion, we have introduced a terrace-step-kink model which
generalizes the Villain-Grempel-Lapujoulade description of vicinal
surface roughening by including a short-range repulsion between
parallel kinks on neighbouring steps as well as on the same step. 
If the strength of this repulsion is sufficient, a VDOF, vicinal
disordered phase, could be stabilized in a temperature window between
the ordered flat and the disordered rough vicinal phases.
 
Recent STM data are speculatively interpreted to suggest that a possible
realization of this new scenario could be found in Ag(115) between 440 K
and 490 K. A refinement of the STM data analysis, particularly concerning
the short-distance correlations between kinks, is very desirable, as it
could now confirm or deny this conjecture.

\vspace{5mm}
\begin{center}
{\bf Acknowledgements}
\end{center}
We thank J. W. M. Frenken for valuable discussions. This work has been
financially supported by INFM/G through a grant for one of us (S.P.),
by INFM PRA LOTUS, and by MURST COFIN97.

\newpage
\begin{center}
{\bf Figure Captions}
\end{center}
\begin{description}
\item[{\bf Fig.\,1 :}]  Schematic view of the TSK vicinal surface,
including excitations and their energy cost. Atomic positions along
the step are all odd or all even, depending on which sublattice
the terrace atoms on the left of the step belong to. The total
energy of the pair of consecutive parallel kinks on the same step
is $2W_1+W_2$, while the energy of the two nearby parallel kinks
on the steps labelled $m$ and $m+1$ is $2W_1+U_2$.

\item[{\bf Fig.\,2 :}]  A snapshot from the Monte Carlo simulation of
the TSK model with $U_2=W_2=+\infty$. Here, the step number is 16 and
$T=\infty$ (only a small part of the lattice with 8 steps is shown).
The viewpoint is from [115], and darker atoms are deeper than lighter
atoms. Left and right kinks strictly alternate along the steps, leading
to an extremely idealized realization of a VDOF surface (i.e., defect
number and amount of correlation between the $u$ values of adjacent
steps both negligible).

\item[{\bf Fig.\,3 :}]  MC results for the TSK model with $U_1=19$ K,
$W_1=1183$ K, $U_2=260$ K, and $W_2=0$ (from 2 to 4 million MC sweeps
were produced at each temperature point). Simulation data are plotted
for three different sizes, $N_x=12$ ($\triangle$), 18 ($\Box$), and 24
($\circ$), with $N_y=60N_x$ in each case. Results are shown for $a)$ the
DOF order parameter $P$, $b)$ its susceptibility $\chi_P$, $c)$ the
specific heat $C/k_B$, and $d)$ the average square lateral step excursion
$\delta u^2$.
In particular, the broad shoulder of $\delta u^2$ at 380 K appears to be
the fingerprint of VPR, while roughening does not occur below 390 K.
The order parameter and susceptibility are consistent with a size-limited
critical transition near 380 K. The heat capacity is non-critical and
size-independent, as expected for negative $\alpha$.
\end{description}
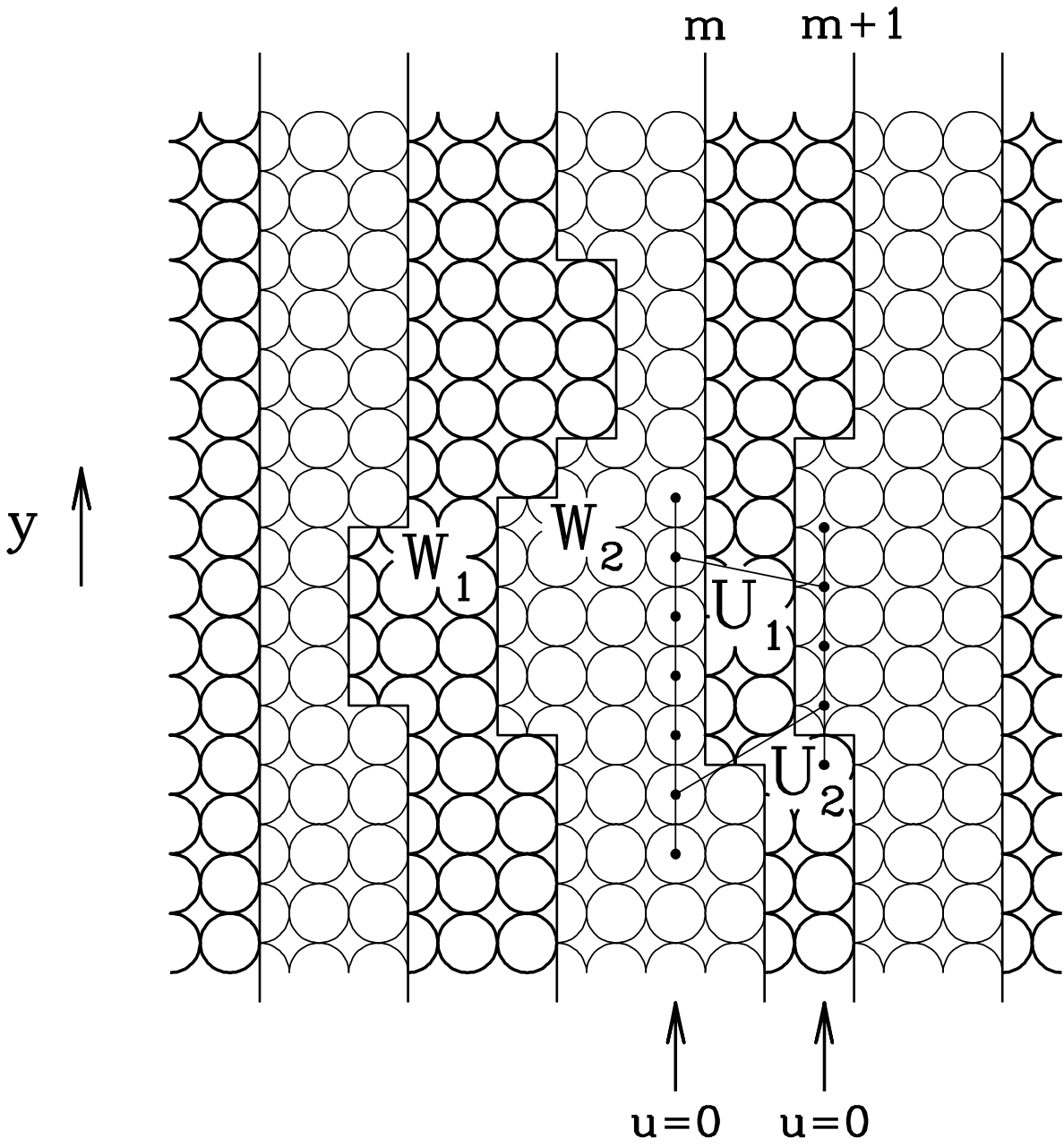
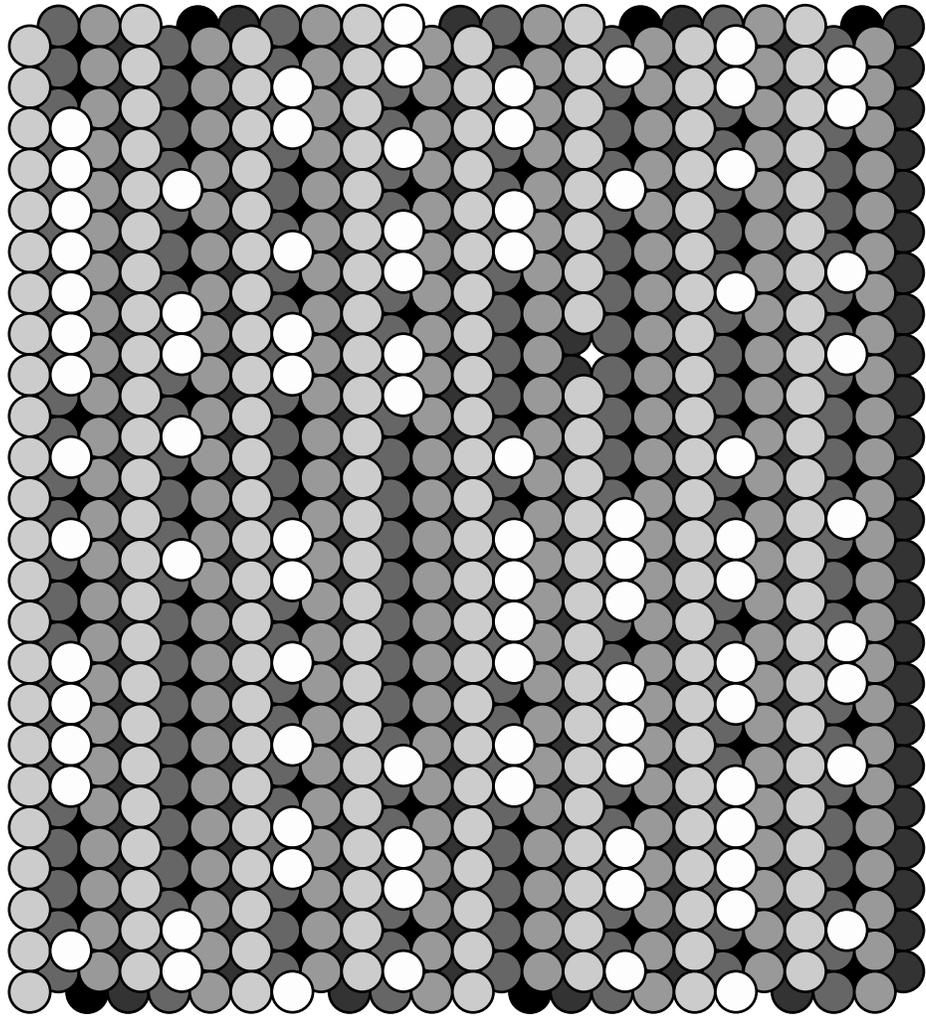
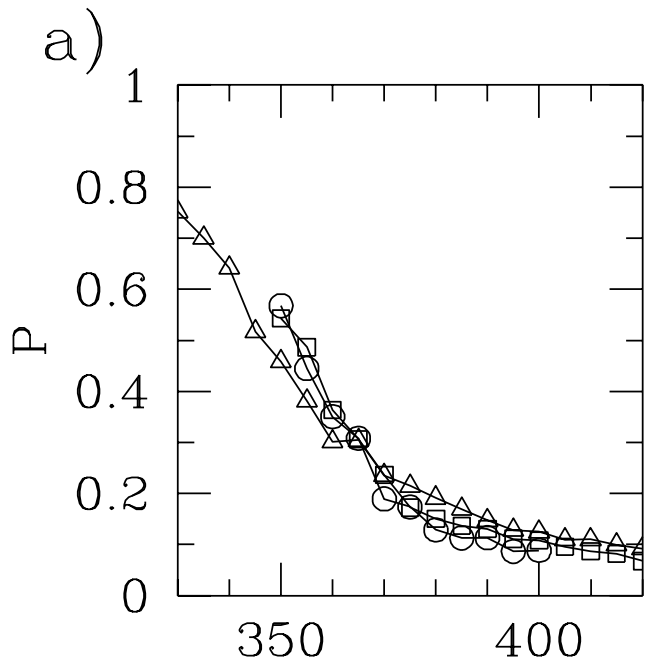
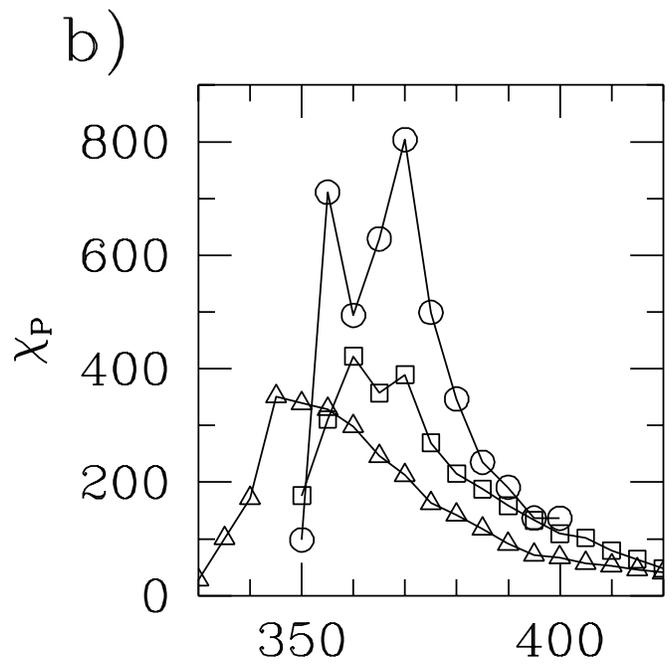
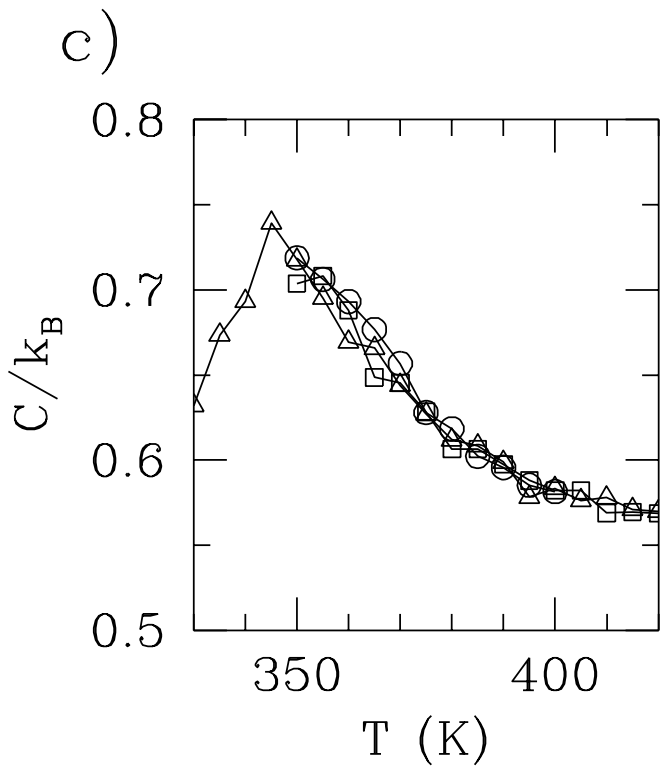
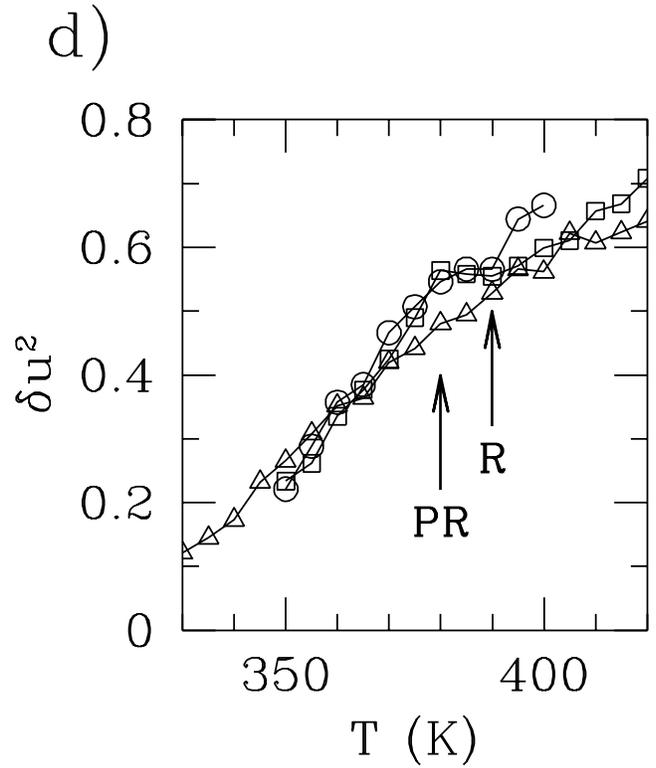
\end{document}